\documentclass[aps,prd,twocolumn,superscriptaddress,preprintnumbers,nofootinbib]{revtex4-1}

\usepackage{amsfonts}
\usepackage{amsmath}
\usepackage{amsthm}
\usepackage{amssymb}
\usepackage{array}
\usepackage{dcolumn}
\usepackage[svgnames]{xcolor}
\usepackage[hidelinks]{hyperref}
\hypersetup{
    colorlinks=true,
    linkcolor=NavyBlue,
    filecolor=NavyBlue,
    urlcolor=NavyBlue,
    citecolor=NavyBlue,
    }

\usepackage{graphics}
\usepackage{tikz}
\usepackage{graphicx}
\usetikzlibrary{positioning}

\usepackage{adjustbox}
\usepackage{gensymb}
\usepackage{breqn}
\usepackage{braket}
\usepackage{enumitem}
\usepackage{gensymb}

\def\be{\begin{equation}}
\def\ee{\end{equation}}
\def\bea{\begin{eqnarray}}
\def\eea{\end{eqnarray}}

\def\gev{\, {\rm GeV}}
\def\mev{\, {\rm MeV}}
\def\kev{\, {\rm keV}}

\def\s{\, {\rm s}}
\def\m{\, {\rm m}}

\newcommand{\gsim}{\lower.7ex\hbox{$\;\stackrel{\textstyle>}{\sim}\;$}}
\newcommand{\lsim}{\lower.7ex\hbox{$\;\stackrel{\textstyle<}{\sim}\;$}}

\newcommand{\cm}{\rm cm}

\begin{document}

\hfill \preprint{ MI-TH-215}

\title{ Low-mass inelastic dark matter direct detection via the Migdal effect}

\author{Nicole F.~Bell}
\email{n.bell@unimelb.edu.au}
\affiliation{ARC Centre of Excellence for Dark Matter Particle Physics$,$ \\~School of Physics$,$~ The~ University~ of~ Melbourne$,$~ Victoria~ 3010$,$~ Australia}

\author{James B.~Dent} 
\email{jbdent@shsu.edu}
\affiliation{Department of Physics$,$~ Sam ~Houston~ State~ University$,$~ Huntsville$,$~ TX~ 77341$,$~ USA}

\author{Bhaskar Dutta}
\email{dutta@physics.tamu.edu}

\author{Sumit~Ghosh}
\email{ghosh@tamu.edu}
\affiliation{Mitchell Institute for Fundamental Physics and Astronomy$,$ Department~ of ~Physics ~ and~ Astronomy$,$\\ Texas A$\&$M University$,$~College~ Station$,$ ~TX ~77843$,$~ USA}

\author{Jason Kumar}
\email{jkumar@hawaii.edu}
\affiliation{Department of Physics and Astronomy$,$~ University~ of~ Hawaii$,$~ Honolulu$,$~ HI~ 96822$,$~ USA}

\author{Jayden L.~Newstead}
\email{jnewstead@unimelb.edu.au}
\affiliation{ARC Centre of Excellence for Dark Matter Particle Physics$,$ \\~School of Physics$,$~ The~ University~ of~ Melbourne$,$~ Victoria~ 3010$,$~ Australia}

\begin{abstract}
We consider searches for the inelastic scattering of low-mass dark matter at direct detection experiments, using the Migdal effect.  We find that there are degeneracies between the dark matter mass and the mass splitting that are difficult to break.  Using XENON1T data we set bounds on a previously unexplored region of the inelastic dark matter parameter space. For the case of exothermic scattering, we find that the Migdal effect allows xenon-based detectors to have sensitivity to dark matter with ${\cal O}(\mev)$ mass, far beyond what can be obtained with nuclear recoils alone.
\end{abstract}

\maketitle

\section{Introduction} \label{sec:introduction}

There has been significant recent interest in methods of detecting low-mass dark matter at direct detection experiments~\cite{Essig:2011nj, Essig:2012yx, Graham:2012su, An:2014twa, Essig:2015cda, Hochberg:2015pha, Derenzo:2016fse, Bloch:2016sjj, Hochberg:2016ntt, Hochberg:2016ajh, Kouvaris:2016afs, Essig:2017kqs, Budnik:2017sbu, Bunting:2017net, Knapen:2017ekk, Hochberg:2017wce, Hertel:2018aal, Dolan:2017xbu, Bringmann:2018cvk, Emken:2019tni, Essig:2019kfe, Ema:2018bih, Bell:2019egg, Trickle:2019ovy,Trickle:2019nya, Griffin:2019mvc, Baxter:2019pnz, Kurinsky:2019pgb, Catena:2019gfa, Griffin:2020lgd, Flambaum:2020xxo}.  
The main difficulty to be overcome is that, for low-mass dark matter, the 
recoil energy deposited in the detector is typically small relative to threshold values needed for detection.  For the case where the dominant interaction is with nucleons, this problem is exacerbated by the 
fact that nuclear recoils are more difficult to detect than electron recoils.
Some particularly promising detection strategies involve new analysis techniques, rather than new detector technologies. A particularly useful 
strategy, which has been the subject of several recent studies, is the Migdal effect~\cite{Migdal:1941, Vergados:2004bm, Bernabei:2007jz, Ibe:2017yqa, Dolan:2017xbu, Bell:2019egg, Essig:2019xkx, Liu:2020pat, GrillidiCortona:2020owp, Dey:2020sai}.   
These studies have largely focused on elastic nuclear scattering.  However, inelastic dark matter scattering (iDM) is a generic feature of many classes of dark matter models~\cite{TuckerSmith:2001hy, TuckerSmith:2004jv, Finkbeiner:2007kk, Arina:2007tm, Chang:2008gd, Cui:2009xq, Fox:2010bu, Lin:2010sb, DeSimone:2010tf, An:2011uq, Pospelov:2013nea,Finkbeiner:2014sja, Dienes:2014via, Barello:2014uda, Bramante:2016rdh,Bell:2018pkk, Jordan:2018gcd,Bell:2020bes}.  Here we discuss the Migdal effect in the context of inelastic dark matter scattering.

If dark matter scatters off a nucleus, then electrons may 
be released via the Migdal effect.  Essentially, the 
electron cloud is boosted relative to the scattered nucleus, 
which can result in electron emission.  Because many direct 
detection experiments are more sensitive to energetic 
electrons than to recoiling nuclei, these Migdal electrons 
can provide the leading channel for the direct detection of low-mass dark matter.

The spectrum of energy deposition in the detector can be altered if dark matter-nucleus scattering is inelastic.  Indeed, the Migdal effect itself can be thought of as a type of inelasticity in the DM-nucleus scattering, as the Migdal electrons carry away energy, but negligible momentum.
Dark matter-nucleus scattering can exhibit inelasticity in two other ways: by exciting a low-lying nuclear state, or by changing the dark matter particle mass. Here we consider the latter case, by assuming the dark matter particle emerging from the scattering process has a different mass than the incoming dark particle.  If the outgoing mass is assumed to be larger (smaller) than the incoming mass, the scattering is said to be endothermic (exothermic).

Although inelastic dark matter scattering has been studied in-depth in the context of models to explain the DAMA excess~\cite{TuckerSmith:2001hy,TuckerSmith:2002af,Chang:2008gd,Finkbeiner:2009ug,Kang:2019uuj}, it is in fact a generic feature of some classes of dark matter models.  While we will not explore particular inelastic dark matter models for the purpose of this paper, we provide the following as an illustrative example. Inelastic scattering mediated by a dark photon with a vector coupling to the dark matter is generic in any model where dark matter is only charged under spontaneously broken continuous symmetries.  The reason is that a gauge boson can only have a vector coupling to a complex degree of freedom.  But if all of the continuous symmetries under which the dark matter is charged are spontaneously broken, then the dark matter is generically expected to split into two real degrees of freedom.  Since one cannot form a vector current with a single real degree of freedom, the dark photon must instead couple to an off-diagonal vector current, yielding inelastic scattering~\cite{Dutta:2019fxn}.  

The relevant parameters of the dark matter model include the mass splitting, $\delta$, as well 
as the mass $m_\chi$ and dark matter-nucleon cross section $\sigma_{\chi n}$.  We will demonstrate 
that there are degeneracies among these parameters that cannot easily be broken with 
the data from direct detection experiments alone.
We will also show that, in the case of exothermic scattering, the Migdal effect provides a unique opportunity to probe very low-mass dark matter.  

The plan of this paper is as follows.  In Section \ref{sec:Migdal}, we derive the electron recoil spectrum arising from the Migdal effect in the case of inelastic dark matter scattering. In Section \ref{sec:eventRate} we explore the shape of the recoil spectrum with some illustrative examples.  In Section \ref{sec:Results} we present our results and comment on the distinguishability of the various scenarios under consideration. In Section \ref{sec:conclusion}, we conclude with a discussion of our results and future avenues.

\section{The Migdal Effect with Inelastic Scattering}
\label{sec:Migdal}

We focus on electron ionization in liquid xenon as a result of DM-nucleus scattering, through the Migdal effect.  As noted in~\cite{Aprile:2019jmx}, bremsstrahlung and electron excitation are expected to be subleading effects in xenon-based detectors.  Similarly, we will follow the isolated-atom approximation (see~\cite{Ibe:2017yqa}), which is expected to be a good approximation for xenon at the relevant momentum transfer.

The non-relativistic DM-atomic scattering process is
$\chi A \rightarrow \chi' A$.  
In the center-of-mass frame, we find
\bea
\frac{1}{2} \mu v^2 &=& E_{\chi'} + E_A +\Delta,
\eea 
where $\mu = m_\chi m_A /(m_\chi + m_A)$ is the reduced 
mass of the initial $\chi-A$ system, $v$ is the relative 
velocity of the two initial particles, and $E_{\chi'}$ and 
$E_A$ are the kinetic energies of the outgoing dark particle 
and the atom, respectively.  
Here, $\Delta = E_{\rm EM} + \delta$ represents the amount of the initial 
kinetic energy lost to inelastic effects, including the excitation of Migdal electrons, $E_{\rm EM}$, and the mass splitting, $\delta$, between 
$\chi$ and $\chi'$.  Note that we have not necessarily 
assumed $m_\chi < m_{\chi'}$; as such, $\delta$ may be 
negative.

In the frame of the detector, we then can express the 
atomic recoil energy as
\bea
E_R &=& \frac{\mu^2}{m_A} v^2 
\left[ 1 - \frac{\Delta}{\mu v^2} 
- \sqrt{1 - \frac{2 \Delta}{\mu v^2} } 
\cos \theta_{cm} \right] ,
\label{eq:ER}
\eea
where $\theta_{cm}$ is the dark matter scattering angle 
in the center-of-mass frame, and where we have made the 
approximation $\mu_f=m_{\chi'} m_A / (m_{\chi'} + m_A) 
\sim \mu$ (for our numerical analysis we have retained the full expression with $\mu_f\neq\mu$). The maximum possible inelastic energy for a given velocity is given by,
\be
\Delta = \frac{1}{2}\mu v^2
\ee
which ensures the argument of the square root remains positive.
 
For velocity-independent spin-independent scattering, the 
scattering matrix element is independent of $\theta_{cm}$.  
If the DM interacts with the nucleus through $s$-channel 
exchange of a mediator with mass $m_\phi$, then the 
squared matrix element depends on $E_R$ as 
$\propto m_\phi^4 / (m_\phi^2 + 2m_A E_R)^2$,
yielding a differential atomic scattering rate which scales 
as

\bea
\frac{d^2 R}{dE_R dv} &\propto& 
\frac{m_\phi^4}{(m_\phi^2 + 2m_A E_R)^2} 
\frac{m_A}{2\mu^2 v^2 } 
[v f(v)] F^2 (q),
\eea
where $f(v)$ is dark matter velocity distribution and 
$F(q)$ is the nuclear form factor.  Henceforth, we will assume 
a contact interaction.
\begin{figure}
    \centering
    \includegraphics[width=0.9\columnwidth]{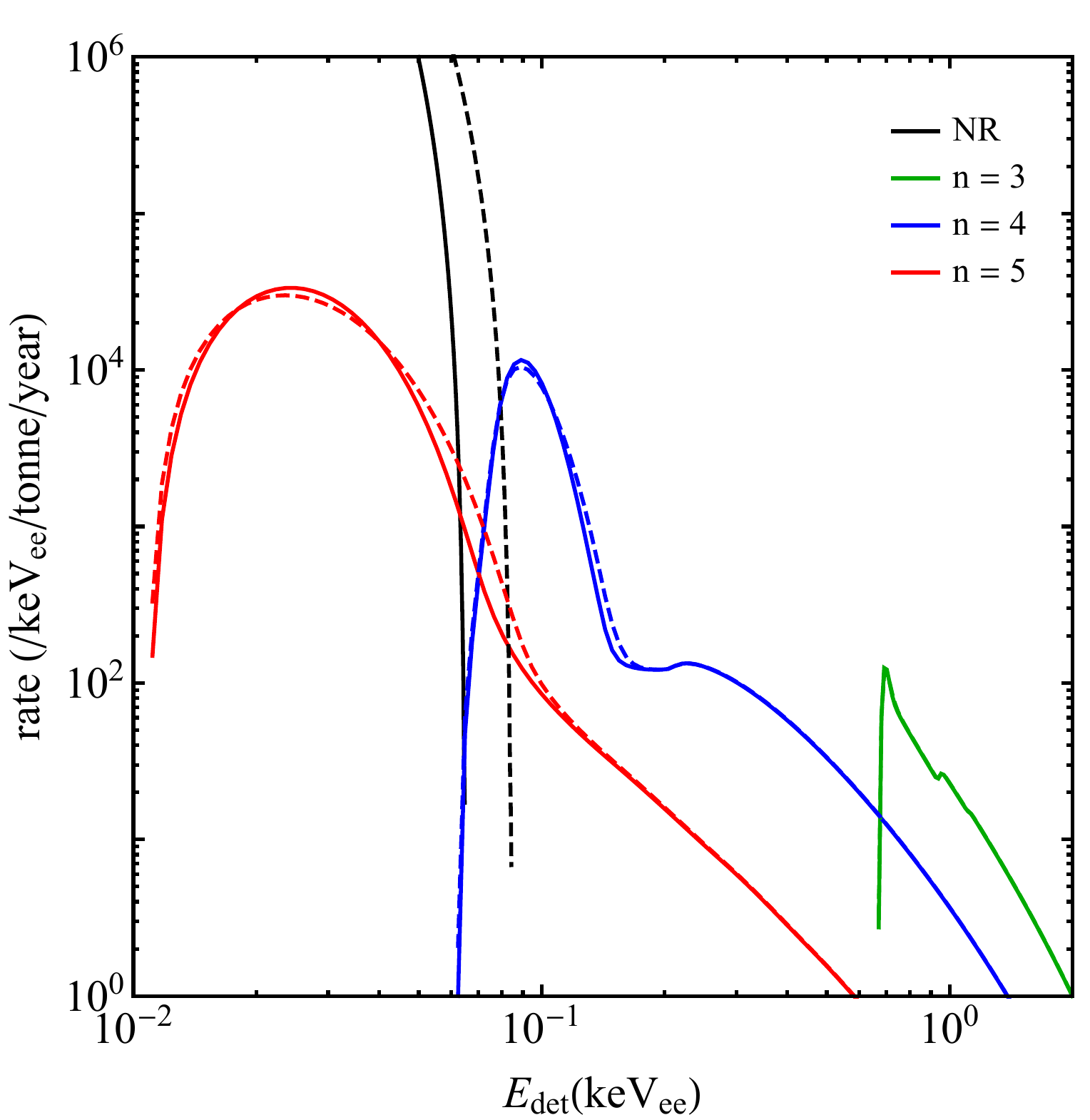}
    \caption{The rate of Migdal events from xenon's $n~=~3,4,5$ shells for elastic DM scattering with a cross section of $10^{-40}$~cm$^2$, comparing a constant quenching factor (solid) with a quenching factor from Lindhard theory (dashed). } 
    \label{fig:migdal_lind}
\end{figure}
The electromagnetic energy spectrum is then given by
\bea
\frac{d^3 R}{dE_{\rm EM} dE_R dv} &=& 
\frac{d^2 R}{dE_R dv} \times \frac{1}{2\pi} \sum_{n,\ell}
\frac{d}{dE_{\rm EM}} p^c_{q_e} (n,\ell \rightarrow E_e), \nonumber\\
\label{eq:migRate}
\eea 
where $p^c_{q_e}$ is the ionization probability (values 
for xenon can be found in~\cite{Ibe:2017yqa}). Here we include the rates for Migdal electrons 
originating from the $n=3,4,5$ shells. To obtain the deposited energy spectrum we must sum the contributions from the nuclear recoil, $E_R$, and the electromagnetic energy, $E_{\rm EM}$, which includes the ejected electron $E_e$ and the atomic de-excitation energy $E_{nl}$:
\be
E_{\mathrm{det}} = \mathcal{L} E_R + E_e + E_{nl}
\ee
where we have included a quenching factor, $\mathcal{L}$, on the nuclear recoil energy. The detected energy spectrum is obtained by integrating Eq.~\ref{eq:migRate} over the atomic recoil energies, incoming WIMP velocities and enforcing energy conservation:
\bea
\frac{dR}{dE_{\rm det} dE_R dv} = \int_{v_{\rm min}}\!\!dv\int dE_R\frac{d^3 R}{dE_{\rm EM} dE_R dv}\nonumber\\\
\times \delta(E_{\rm det} - \mathcal{L} E_R - E_e - E_{nl}). \\
\nonumber
\eea 
The minimum and maximum atomic recoil energy limits are obtained from the Eq.~\ref{eq:ER} with $\theta\rightarrow0$ and $\pi$, respectively.
We assume the WIMP speed distribution to be Maxwell-Boltzmann with $v_0=220$ km/s, cut off at the local escape velocity $v_{\rm esc} = 544$ km/s. With the $\mu_i=\mu_f$ simplification, the minimum WIMP velocity is approximately given by:
\be
v_{\rm min} = \frac{|E_R m_T + \mu\Delta|}{\sqrt{2E_R m_T}\mu}.
\ee
In Migdal calculations it has become customary to take a constant value of $\mathcal{L}=0.15$, which introduces an error that is sub-dominant compared to the uncertainty in the atomic ionization probabilities. To illustrate this we first calculate the rate of the Migdal effect in xenon using both a constant quenching factor and using a quenching factor from Lindhard theory as implemented in~\cite{Lenardo:2014cva}. The result is shown in Fig.~\ref{fig:migdal_lind}. The effect on the differential rate is small and the effect on the total rate is negligible.

As noted in \cite{Knapen:2020aky}, the impulse approximation used in the calculation of the ionization probabilities will break down if the collision time scale (set by the inverse of the recoil energy $E_R^{-1}$) is greater than the time taken for the atom to traverse its potential (set by the inverse of the phonon frequency $\omega_{ph}^{-1}$). In a fluid we will take this cutoff as the time taken for an atom to traverse the average inter-atomic distance at the sound speed. For xenon at 170K the time scale is $t_1=a/v_s\approx 10^{-12}\s$, therefore we conservatively set the cutoff recoil energy to be: $E_R>100t_1^{-1}\approx50$ meV. This limit has an increasingly large effect at lower dark matter mass and target thresholds. Sufficiently small dark matter masses will have a maximum recoil energy that falls below this cutoff, therefore there is a minimum mass for which the impulse approximation is valid.  For elastic scattering we restrict ourselves to $m_\chi\gsim 0.02$~GeV. For exothermic (endothermic) scattering the limits are dependent on the mass splitting, for example at $-10$~keV (10~keV) we set $m_\chi\gsim 0.6$~MeV ($0.4$~GeV).

\section{The Migdal event rate and its degeneracies}
\label{sec:eventRate}
\begin{figure}
    \centering
    \includegraphics[width=0.9\columnwidth]{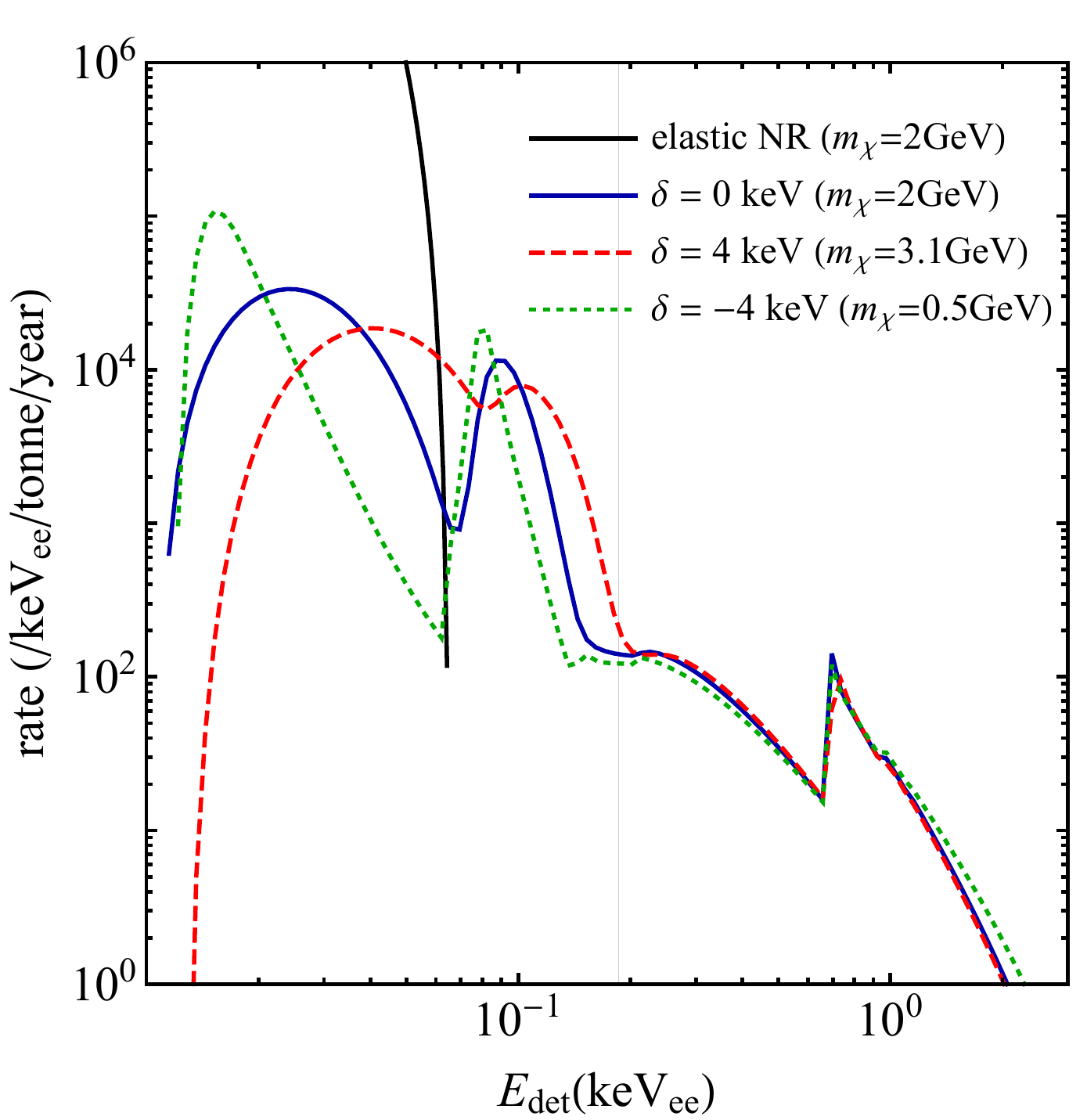}
    \caption{Rate of Migdal events for elastic (blue) and inelastic endothermic (red) and exothermic (green) DM scattering, for parameters that illustrate the $m_\chi - \delta$ degeneracy. Also shown is the elastic nuclear recoil rate (black). The vertical gray line corresponds to the threshold of the XENON1T S2-only analysis. The cross sections of the elastic, exothermic and endothermic curves are (1, 0.68, 3.5)$\times 10^{-40}$ cm$^2$, respectively.} 
    \label{fig:migdal_m2}
\end{figure}

\begin{figure}
    \centering
    \includegraphics[width=0.9\columnwidth]{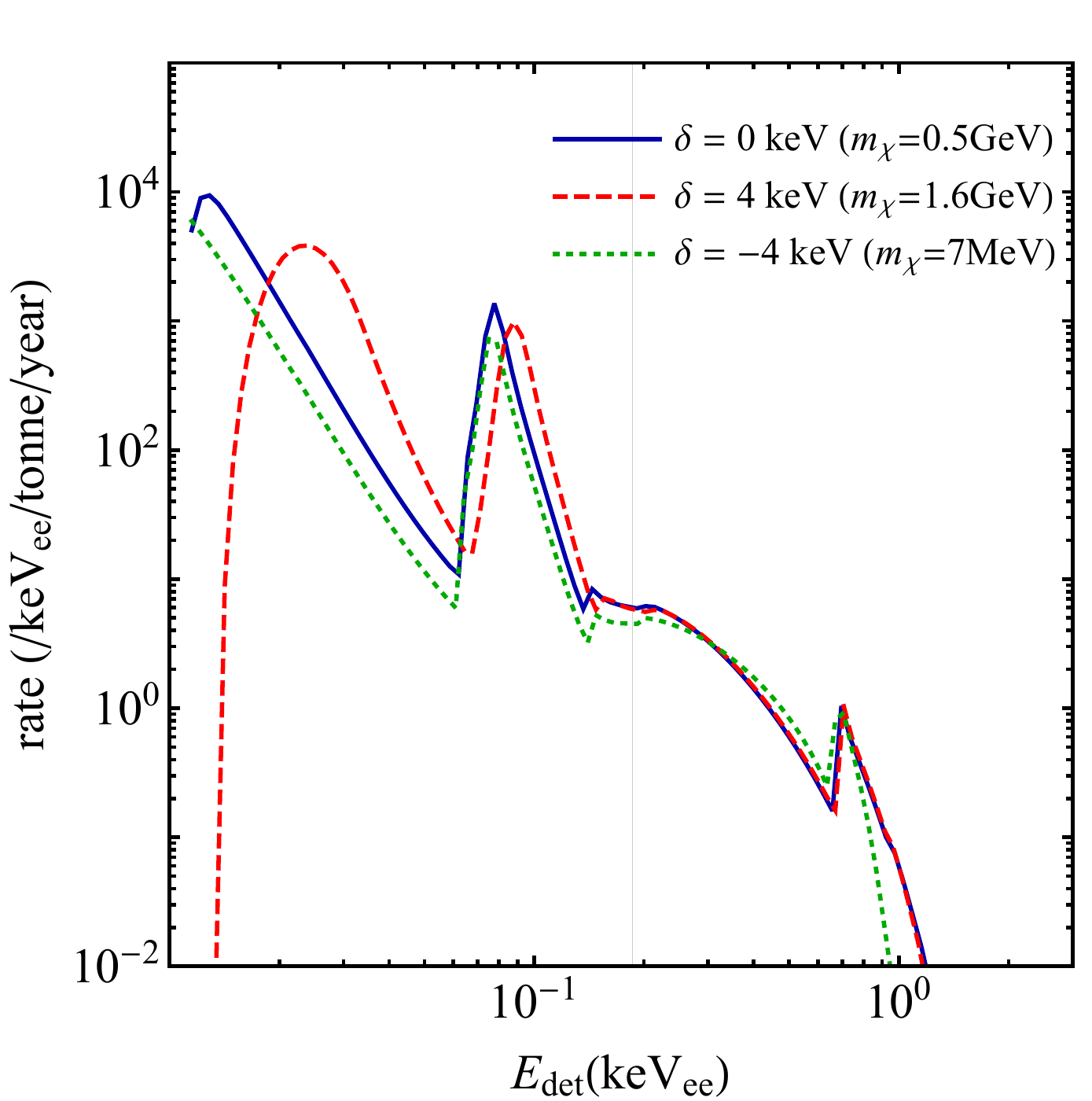}
    \caption{Rate of Migdal events for elastic (blue), endothermic (red) and exothermic (green) DM scattering. The vertical gray line corresponds to the threshold of the XENON1T S2-only analysis.  The cross sections of the elastic, exothermic and endothermic curves are (1, 0.0084, 27)$\times 10^{-40}$ cm$^2$, respectively.}
    \label{fig:migdal_mp5}
\end{figure}

In Fig.~\ref{fig:migdal_m2} we plot the differential 
event rate of nuclear (solid black) and Migdal (solid blue) events as a 
function of the detected energy (in ${\rm keV_{ee}}$), 
for $m_{\chi} = 2~\gev$ and $\sigma_{SI} = 10^{-40} ~\cm^2$.  Note, the nuclear recoil energies would be below threshold for XENON1T. For comparison, in the exothermic case we have assumed that all dark matter is in the excited state (see the following section for discussion of this point). 

The shape of the Migdal electron recoil spectrum can be largely characterized by 3 key features: by the minimum energy (below which no events are seen), the maximum recoil energy, and the peak of the spectrum.
The minimum energy is determined by the energy released 
when a higher-shell electron de-excites to fill the orbital vacated 
by a Migdal electron.  This minimum energy is thus determined by the 
energy levels of the target nucleus, and is largely 
independent of the dark matter parameters.

The maximum electromagnetic energy is given by 
\bea
\label{eq:Emax}
E_{\rm EM}^{ max} &=& \frac{1}{2}\mu v_{max}^2 - \delta ,
\nonumber\\
&\sim& \frac{1}{2}m_{\chi} v_{max}^2 - \delta.
\label{eqn:degen}
\eea
where $v_{max}$ is the maximum DM-nucleus relative 
speed, and we have assumed $m_\chi \ll m_A$.  This 
amounts to the situation in which, in center-of-mass 
frame, all of the kinetic energy is absorbed by the 
various forms of inelasticity (the boost to laboratory 
frame shifts $E_{\rm EM}^{ max}$ by a negligible amount).  
We see that this maximum is independent of the target material, 
and of the initial orbital of the Migdal electron.  
If $v_{max}= v_{\rm esc} + v_{\rm Earth}\sim 800$~km/s  
is taken to be the maximum speed with 
respect to the Earth of a particle at the galactic 
escape speed, then there is a degeneracy between 
$m_\chi$ and $\delta$ in the determination of 
$E_{\rm EM}^{ max}$. In practice, the probability of a Migdal electron falls 
with both the atomic recoil energy and the electron energy and thus
the rate falls steeply toward $E_{\rm EM}^{ max}$.

The peak of the Migdal electron spectrum (for any shell) occurs just above 
the minimum energy, with an amplitude which is proportional 
to the DM-nucleus scattering cross section.  
We thus see that there should be a one-parameter 
family of models for which the Migdal electron energy spectrum is nearly identical.  
In Figure~~\ref{fig:migdal_m2}, we 
illustrate the approximate degeneracy between the mass and the mass splitting by plotting best-fit scenarios 
with $\delta =-4~\kev$ and $\delta = 4~\kev$. Note that the shift in the mass of 
the incoming particle is roughly what one would expect from Eq.~\ref{eqn:degen}.
In Figure~\ref{fig:migdal_mp5} we make 
a similar plot with the elastic case being $m_\chi = 0.5~\gev$. The best-fit rate is found by minimizing the $\chi^2$ of the integrated rate in 5 log-spaced bins above the XENON1T threshold, i.e. in the range $E_{\mathrm{det}}=0.186-2~\kev$ ($E_{\mathrm{det}}=0.186-1.5~\kev$ for the $m_\chi = 0.5~\gev$ case). We find that while
the rates are almost degenerate above threshold, we see that a opportunity to resolve the
degeneracy exists below threshold. Therefore, future experiments with $\mathcal{O}(10$ eV) thresholds will not suffer greatly from this degeneracy.

\section{Bounds and Sensitivities}
\label{sec:Results}

The event rate at a direct detection experiment depends 
on both the dark matter-nucleus inelastic scattering cross 
section ($\chi_i A \rightarrow \chi_j A$), 
and the fraction of dark matter particles which 
are in the lighter (heavier) state, in the case of 
endothermic (exothermic) scattering.  We denote this 
fraction as $f_\ast$.  As usual, we report a normalized-to-nucleon 
scattering cross section, in order to facilitate the 
comparison of sensitivities of different detectors with 
different target materials.  But some care must be taken 
in the case of inelastic scattering; for some values of 
mass splitting, inelastic scattering with some targets 
is kinematically allowed on Earth, while scattering with 
a proton would be kinematically forbidden.  But since 
we consider scenarios with $|\delta| \ll m_\chi$, the 
scattering matrix element is largely independent of 
$\delta$, with the dependence of the cross section on 
$\delta$ arising from the phase space factors.  Thus, 
we report the cross section $\sigma_{\chi p}$, which 
is the dark matter-proton scattering cross section 
extrapolated to $\delta = 0$; given this quantity, 
one can 
determine the dark matter-nucleus inelastic scattering 
cross section  for the 
given $\delta$ and for any choice of nuclear target $A$.

Dual phase liquid noble detectors have two detection channels: primary scintillation light (S1) and delayed proportional scintillation (S2) which is produced by primary ionization that is drifted through the liquid and accelerated in the gas phase. The detection of S1 and S2 signals allows the reconstruction of the 3D position, energy deposition and nuclear/electronic recoil discrimination. The most sensitive low-energy threshold analysis comes from looking at the ionization channel alone. We set upper limits on the quantity $f_\ast \sigma_{\chi p}$ using the S2-only data set from the XENON1T experiment~\cite{Aprile:2019xxb}. This single channel analysis does not discriminate nuclear and electronic recoils and thus it can be used to place bounds on the cross section from both nuclear recoils and the Migdal effect. We perform a single bin analysis by integrating the total event rate in the range $E=0.186-3.8$ keV$_{ee}$, taking into account the energy dependent efficiency. The total number of events observed in the 22 tonne-day exposure was 61, while 23.4 were expected from background. Using this data we can place an upper limit (at 90\% confidence) of 49 expected events coming from dark matter. We use this upper limit to place bounds on
$f_\ast \sigma_{\chi p}$ for both nuclear recoil events and Migdal events.

\begin{figure}
   \centering
    \includegraphics[width=0.9\columnwidth]{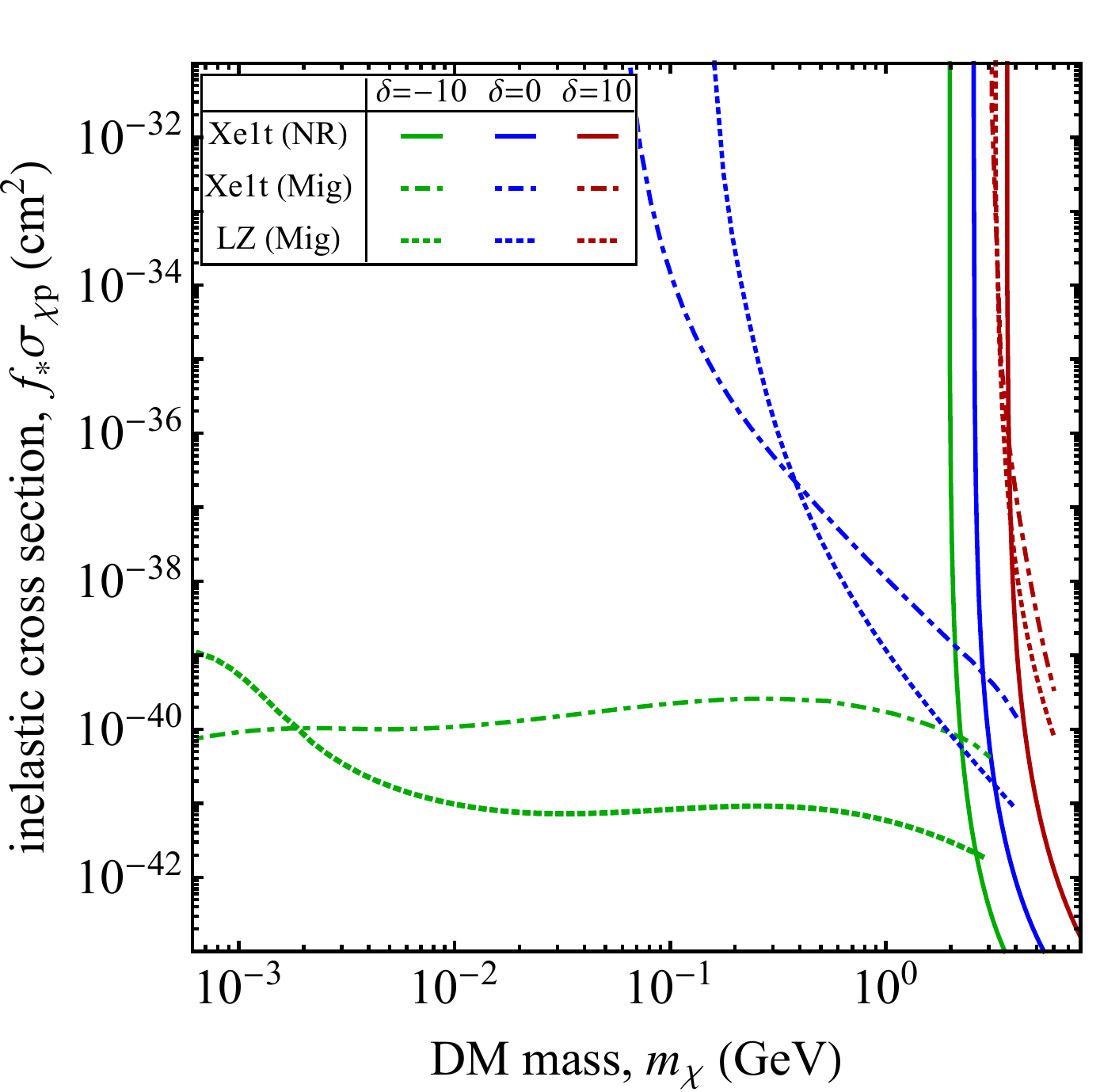}
        \caption{Bounds on the inelastic cross section from XENON1T on nuclear recoils (solid), XENON1T with the Migdal effect (dashed) and the projected LZ sensitivity with the Migdal effect (dotted). 
    These are each plotted for for exothermic (green), elastic (blue) and endothermic (red) interactions, where $\delta= -10, 0, 10 \kev$ respectively. We restrict our bounds to $m_\chi \ge 0.6$ MeV, where we can be sure the impulse approximation is valid.}
    \label{fig:mig_bounds_xe1t}
\end{figure}

Additionally, we project the sensitivity of LZ~\cite{Akerib:2019fml, Akerib:2021qbs} assuming a 1000 day exposure of the 5.6 tonne volume. Using a region of interest of $E=0.5-4$ keV$_{ee}$ (assuming XENON1T's energy dependent efficiency from~\cite{Aprile:2020tmw}), an expected background of $2\times 10^{-5}$/kg/day/keV \cite{Akerib:2021qbs} and background uncertainty of 15\%, we place an upper limit of 80 expected events from dark matter. The resulting bounds are shown in Fig.~\ref{fig:mig_bounds_xe1t} for elastic, exothermic, and endothermic interactions. The exo/endothermic cases are calculated for a benchmark mass-splitting of $\pm 10\kev$. These mass splittings demonstrate the extremal behaviour of the Migdal bounds in the inelastic parameter space, from becoming essentially mass independent, to becoming no better than the nuclear recoil bounds. 

For nuclear scattering,  exo/endothermic inelasticity 
only shifts the bounds by $\sim 1$ GeV.
As mentioned previously, the maximum energy that can go into the Migdal electrons is approximately proportional to the dark matter kinetic energy, less the mass-splitting energy (Eq.~\ref{eqn:degen}). This drives the loss in sensitivity of the elastic and endothermic interaction curves, while the exothermic rate remains above threshold for very small dark matter masses.  Essentially, even for very small dark matter mass and kinetic energy, the energy yielded by exothermic scattering is sufficient 
to yield Migdal electrons above threshold. For very small $m_{\chi}$, we estimate that LZ's sensitivity will not exceed limits from XENON1T.  This is due to our estimate for LZ's threshold, 
which is larger than that of XENON1T's S2-only analysis.\footnote{We note that LZ may also perform an S2-only analysis, improving their low-mass sensitivity. However, the achievable threshold, background and exposure of such an analysis is unknown.}

It is interesting to note that, for $m_\chi \ll \gev$, XENON1T is not able to detect exothermic nuclear recoils even for $\delta = -10\kev$. Even though a significant amount of energy is released in exothermic scattering, for such light dark matter, the vast majority of this energy is transferred to the outgoing DM or the Migdal electrons, but not to nuclear recoils.

It is worthwhile to estimate the range of $f_\ast$ which are 
reasonable in the case of exothermic scattering.  If the 
decay $\chi \rightarrow \chi' \gamma$ can proceed rapidly 
through a magnetic dipole interaction, then $f_\ast$ would be 
expected to be very small.  But if the coefficient of this 
operator is negligible, then the leading decay process 
would be $\chi \rightarrow \chi' \gamma \gamma \gamma$, 
with a rate which could be long compared to the age of the 
Universe~\cite{Dienes:2017ylr}.  In this case, the dominant 
process leading to the depletion of $\chi$ is exothermic 
scattering in the early Universe, between the temperature at 
which the dark matter decouples from chemical equilibrium with 
the Standard Model and the temperature  at which the 
$\chi$ and $\chi'$ chemically decouple from each other. 

We consider $|\delta| \leq 10\kev$, 
therefore the Boltzmann suppression of the heavier state can only be 
relevant at temperatures $T < \delta \ll m_e$, at which point the 
electron-positron number density will already be suppressed. 
We thus find that the dominant process contributing to the 
depletion of the heavier species is $\chi \chi 
\leftrightarrow \chi' \chi'$.  The rate for this process is 
model-dependent, and cannot directly be determined from 
$\sigma_{\chi p}$.  A variety of models were considered in 
Ref.~\cite{Baryakhtar:2020rwy}, with values of $f_\ast$ found in 
the range ${\cal O}(10^{-9} - 10^{-1})$ which depend on $\m_\chi$ and $\sigma_{\chi p}$. However it is possible to realize dark matter scenarios with $m_\chi \sim 1 \mev$, 
$\sigma_{\chi p} \sim {\cal O}(10^{-40}\cm^2)$ in which 
the process $\chi \chi \leftrightarrow \chi' \chi'$ decouples 
at temperatures $\gg \delta$.  In these scenarios, $\chi$ and 
$\chi'$ are cosmologically stable and chemically decoupled at
temperatures for which the mass splitting is irrelevant,
leading to $f_\ast \sim \mathcal{O}(1)$ (see for example secluded DM~\cite{Bramante:2020zos}).

\section{conclusion} \label{sec:conclusion}

Inelastic nuclear scattering is a feature which arises in many classes of dark matter models. We have considered the impact of this feature on searches for the scattering of low-mass dark matter using the Migdal effect. We have found that
there is an irreducible degeneracy in the Migdal electron energy spectrum between DM mass and the inelastic splitting.
The direct detection data alone is thus insufficient to reconstruct the dark matter model.  We also find that, in the case of exothermic scattering, the Migdal effect 
provides sensitivity to dark matter masses as low as 1 MeV, providing a new 
approach to searching for very low-mass inelastic dark sector models.  Although we have focused on xenon-based detectors, these results should generalize to other materials. 

Other than reducing the threshold, one potential way to break the degeneracy between mass and mass splitting is to search for photons produced by decay of the heavier dark state. This may occur at a location displaced from the original scatter, which itself need not even be within the detector volume. The energy of these decay photons would determine the mass splitting, breaking the model degeneracy. Future work could explore the regions of model space where such a detection would be feasible.

\begin{acknowledgments}

The work of BD and SG are supported in part by the DOE Grant No. DE-SC0010813. The work of JK is supported in part by DOE grant DE-SC0010504. NFB and JLN are supported in part by the Australian Research Council. JBD acknowledges support from the National Science Foundation under Grant No. NSF PHY-1820801.

\end{acknowledgments}

\bibliographystyle{apsrev4-1.bst}
\bibliography{migdal.bib}

\end{document}